\documentclass[aip,jcp,amsmath,amssymb,twocolumn,reprint]{revtex4-2}
\setcitestyle{numbers,square,comma}
\usepackage{natbib}
\usepackage{graphicx, color, soul}
\usepackage[normalem]{ulem}

\DeclareMathOperator{\diag}{diag}

\newcommand{\be}{\begin{equation}}
\newcommand{\ee}{\end{equation}}
\newcommand{\bea}{\begin{eqnarray}}
\newcommand{\eea}{\end{eqnarray}}
\newcommand{\bvec}[1]{\mathbf{#1}}
\newcommand{\transp}{^{\top}}

\newcommand{\Eq}[1]{Eq.~\eqref{#1}}
\newcommand{\Eqs}[1]{Eqs.~\eqref{#1}}
\newcommand{\Fig}{Fig. \ref}

\newcommand{\br}{\bvec{r}}

\newcommand{\bR}{\bvec{R}}

\newcommand{\bK}{\bvec{K}}

\newcommand{\bU}{\bvec{U}}

\newcommand{\bM}{\bvec{M}}

\newcommand{\bq}{\bvec{q}}

\newcommand{\HO}{\text{(H}_2\text{O)}}

\def\<{\left\langle}                 
\def\>{\right\rangle}                 

\begin{document}
\title{A computationally efficient subspace harmonic relaxation
  algorithm for coarse-graining of molecular systems with nearly exact thermodynamic consistency}  
\author{João V. M. Pimentel}
\email{joaop@uci.edu}
\author{Vladimir A. Mandelshtam}
\affiliation{Department of Chemistry,
University of California, Irvine, CA 92697, USA}
\begin{abstract}
In a recent paper, J. Chem. Phys. 162, 214101 (2025), a novel approach for the rigidification of a molecular cluster was proposed, in which starting with an all-atom (AA) potential, a coarse-grained (CG) potential for the associated cluster of rigid monomers was constructed directly. The method is based on using the harmonic approximation for the fast intramolecular degrees of freedom. While conceptually primitive, the resulting CG model turned out to be surprisingly accurate for selected water and ammonia clusters. However, as originally formulated, a single evaluation of the CG potential turned out to be much more expensive than the evaluation of the AA potential, since the former required a subspace minimization followed by a subspace normal mode calculation. In this communication, we formulate the approach more broadly, making it applicable, e.g., to coarse-graining a large protein. We also introduce key algorithmic improvements, reducing the cost of the subspace minimization and normal mode calculation. Combined with the fact that the CG simulation requires roughly an order of magnitude fewer Monte Carlo steps to reach similar statistical accuracy for selected observables compared to the AA model, the overall computational cost becomes comparable. These improvements are demonstrated on a water cluster.


\end{abstract}
\maketitle

\section*{Introduction}
\label{introduction}
CG models \cite{Voth2022,Shell2008,Voth2005,Noid2008} are standard tools to reduce the dimensionality of complex systems (e.g., macromolecules and molecular clusters). In CG models such as MARTINI \cite{Marrink2007}, molecular fragments are typically represented by a small number of interaction sites, which constrains internal flexibility compared to AA models. Elastic network models \cite{Tirion1996} further simplify the representation by replacing AA interactions with harmonic restraints between atoms or groups, allowing efficient modeling of low-frequency modes. More recently, related harmonic approximations have been employed to coarse-grain intra-site degrees of freedom, for example through a normal mode treatment of liquid water \cite{Na2023} or a local harmonic approximation to the exact CG potential \cite{Kidder2024}.

Rigid-body CG models are used routinely when one can identify the fast intramolecular degrees of freedom.  The subspace harmonic relaxation (SHR) method, introduced in Ref.~\onlinecite{Pimentel2025} (hereafter referred to as ``Paper I''), where it was originally termed Rigid-HA, goes beyond standard rigid-body treatments by minimizing the potential energy along these fast modes. In contrast, most existing approaches freeze these modes entirely, as done, e.g., in Refs. \cite{Muniz2021,Wales2013}. This allows the construction of a CG potential from the AA potential through constrained minimization combined with harmonic free energy corrections. While nearly exact, the original approach from Paper I, if used directly, remains significantly more expensive than conventional AA simulations, specifically, due to the need to perform at every step a subspace minimization followed by normal mode analysis.

In this paper, we address both numerical bottlenecks. First, we replace the original conjugate gradient optimization with a fixed number ($P = 1$ or 2) of Newton–Raphson iterations, which accelerates convergence within the constrained subspace. Second, we approximate the Hessian by retaining only block-diagonal components corresponding to each ``rigid'' body. This reduces computational cost in both the Newton–Raphson iterations and the normal mode analysis, while still allowing for an accurate constrained relaxation. The approach is in the spirit of block-based approximations used in low-frequency mode analysis of macromolecules \cite{Tama2000}, enabling efficient evaluation without constructing the full Hessian.

Assuming pairwise interactions between $N$ atoms (which is the case, e.g., for water modeled with the q-TIP4P/F AA potential from Ref.~\onlinecite{habershon2009}), the cost of evaluating the potential energy (and its gradient or Hessian) scales as $\sim N^2$ in the worst case, i.e., when all the long range pair interactions are included. In Paper I, we diagonalized the full Hessian matrix of size $3N \times 3N$, which scales as $\sim (3N)^3$. Here, with the aforementioned block-diagonal approximation, the cost of diagonalizing the Hessian is reduced to $\sim \sum_{i=1}^I  (3 N_i )^3$, where $I$ is the number of groups of atoms, each containing $N_i$ atoms. With this simplification, the cost of evaluating the CG potential boils down to that of computing the potential, the gradient, and the block-diagonal Hessian.
Most importantly, these modifications preserve the accuracy and theoretical foundations of the original method, which we demonstrate numerically by applying it to the $\HO_{10}$ cluster. 

It is possible to reintroduce atomic resolution from CG configurations via backmapping, often using machine-learning approaches. For example, Maier and Jackson \cite{Maier2022} predicted quantum observables from CG configurations, while Stieffenhofer et al. \cite{Stieffenhofer2020} sampled AA configurations conditioned on CG structures and compared observables with the original AA ensemble. Wang et al. \cite{Wang2022} proposed a generative scheme for reconstructing the AA Boltzmann distribution from CG configurations. Similarly, Li et al. \cite{Li2020} employ generative models to reconstruct AA structures from CG configurations, and Shmilovich et al. \cite{Shmilovich2022} ensure temporal coherence in backmapped AA trajectories. Backmapping can also be achieved without machine learning: Waltmann et al. \cite{Waltmann2025} use a diffusion-based method to generate AA structures constrained to CG configurations.
 Our method is complementary: by focusing on systems with a clear separation between fast and slow degrees of freedom, we can use a harmonic approximation to reconstruct all-atom configurations nearly exactly. This allows almost exact computation of AA observables and distributions with minimal computational cost, whereas the more general backmapping approaches must handle the full complexity of the missing degrees of freedom and therefore have limited accuracy in practice.

\section*{The minimum energy manifold from Singular Value Decomposition}

Given an $N$-atom molecular system with the AA potential $V(\br)$ ($\br\in\mathbb{R}^{3N}$), we assume that there exists a natural partitioning of the AA space into a CG subspace described by slow degrees of freedom and the subspace (loosely referred to as the intramolecular subspace) described by the fast degrees of freedom. 
More specifically, we are considering a system, which is decomposed into $I$ groups of atoms, each
containing $N_i$ atoms:
\begin{equation}\label{eq:br}
    \br := \left( \br_1,  \dots, \br_I \right),   \ \ \ \ \br_i\in\mathbb{R}^{3N_i}
    \end{equation}
The positions and orientations of these groups are described by the CG coordinates $\bR:=( \bR_1,  \dots, \bR_I )$, which can generally undergo a large amplitude motion; the atoms within each group are assumed to
undergo only low-amplitude vibrations. 
Each such group could be a molecular monomer, described by $L_i=3N_i-6$ intramolecular degrees of freedom $\bq_i\in\mathbb{R}^{L_i}$ , or a part of a larger molecule with $3N_i> L_i>3N_i-6$. Note here that the fact that the $i$th group is connected to another group in the same molecule introduces at least one constraint. 
The total number of ``fast'' or ``stiff'' degrees of freedom is then $L=\sum_{i=1}^I L_i$ and the total number of CG degrees of freedom is $L_{\rm CG}=3N-L$.

Given an AA configuration $\br$, decomposed according to \Eq{eq:br}, consider the mass-scaled Hessian matrix truncated to the block-diagonal form:
\begin{equation}\label{eq:bK}
 \bK=\diag \{ \bK_1, \dots ,\bK_I\} 
\end{equation}
where the $i$th block corresponds to the $i$th group
 \begin{equation} 
\bK_i = \bM_i^{-1/2}\left(\partial^2 V/\partial \br_i^2\right)\bM_i^{-1/2}\in \mathbb{R}^{3N_i \times 3N_i};
\end{equation}
$\bM_i$ is the diagonal mass matrix for the group $i$.
Now, using its $L_i$ largest eigenvalues (or the singular values), $\lambda^i_l$,
and the corresponding orthonormal eigenvectors, $\bU^i_l\in\mathbb{R}^{3N_i}$, construct
the pseudoinverse of the $i$th Hessian block:
\begin{equation} 
\tilde{\bK}_i:=\sum_{l=1}^{L_i} \frac 1 {\lambda^i_l}\bU^i_l \left[\bU^i_l\right]\transp
\end{equation}
Also, for the full Hessian's pseudoinverse, we have
\begin{equation} 
\tilde{\bK}=\mbox{diag} \{\tilde{\bK}_1,\dots,\tilde{\bK}_I\}
\end{equation}

Finally, define the $L_\text{CG}=(3N-L)$-dimensional minimum energy manifold (MEM) ${\cal R^*}\subset \mathbb{R}^{3N}$, which consists of all points $\br^*$ from the full configuration space of the AA system satisfying
\begin{equation} \label{eq:r*}
\tilde{\bK}(\br^*)\ \bM^{-1/2}\nabla V(\br^*)=\mathbf{0},
\end{equation}
where $\bM=\mbox{diag} \{\bM_1,\dots,\bM_I\}$.
In other words, these are the configurations where the system is at equilibrium along the fast degrees of freedom. Note that this definition of MEM is unique, once the atoms have been grouped 
and the group ranks, $L_i$, have been set. At the same time, ${\cal R^*}$ is 
invariant with respect to the choice of coordinates, which makes it, at least formally, different 
from that of Paper I.

\subsection*{Coarse-Grained potential}

In the spirit of Paper I, we can express the AA partition function 
and obtain related expressions for the observables using
\begin{equation}  \label{eq:ZT}
Z(T) = \int_{\cal R^*} d\br^* \exp\left[-\beta F(\br^*;T)\right],
\end{equation}
where the CG potential or, more precisely, the CG free-energy from the excluded intramolecular subspace is estimated using the local harmonic approximation
\begin{eqnarray} \label{eq:Fh}
F(\br^*;T)  &\approx&  V (\br^*)  + F_h(\br^*;T) \ \ \ \ \ \ \mbox{(Quantum)} \\\nonumber
&=& V (\br^*) + \sum_{l=1}^{L} \left\{ \frac {\hbar\omega_l} 2 +
  \frac {\log(1-e^{-\beta\hbar\omega_l})} {\beta} \right\}
\end{eqnarray}
where $\omega_l=\sqrt{\lambda_{l}}$  ($l=1,...,L$) are all the selected normal mode frequencies combined;
$\beta=1/(k_{\rm B}T)$. 
For the harmonic contribution to the free energy, in the classical limit (high temperature, $\beta \hbar \omega_l \ll 1$) one could also use the expression:
\begin{equation} \label{eq:Fh-classical}
F(\br^*;T)\approx V (\br^*) + \frac 1{\beta} \sum_{l=1}^L \log \left(
 \beta\hbar\omega_l\right)  \ \ \ \ \ \ \mbox{(Classical)}
\end{equation}

Regardless of whether or not the nuclear quantum effects are taken into account, assuming 
that one can efficiently sample the MEM and evaluate the integral over $d\br^*$, \Eqs{eq:Fh} and (\ref{eq:Fh-classical}) have an advantage
over the usual formula (involving Monte Carlo integration over the full AA configuration space) due to the elimination of the fast degrees of freedom, and also
due to a potentially significant reduction of the dimensionality of the manifold to be sampled.

\section*{Numerically efficient sampling of the Minimum Energy Manifold: Subspace Harmonic Relaxation algorithm}

While a direct sampling of the MEM is virtually impossible, or at least non-trivial, we present an indirect algorithm that we call the ``Subspace Harmonic Relaxation'' (SHR), in which the sampling is performed using a manifold that closely resembles the MEM.

\subsection*{Preliminary Setup}

Following Paper I, for a given AA configuration of monomer $i$, we define a bijection
\begin{equation} \label{eq:bijection}
\br_i \mapsto (\bR_i,\bq_i)\ \ \   (i=1,...,I),
\end{equation}
in which the vector $\bq_i\in\mathbb{R}^{L_i}$ describes the fast degrees of freedom for monomer $i$ using the monomer frame, and the CG vector $\bR_i\in\mathbb{R}^{3N_i-L_i}$ defines its position and orientation using the laboratory frame (see Paper I for details). 
Also define $\bq_i^{\rm (0)}$  corresponding to the potential energy minimum with respect to
$\bq_i$ for isolated monomer $i$. For the whole system we have  
$\br^{(0)}:=(\br_1^{(0)},\dots,\br_I^{(0)})$ which maps bijectively to $(\bR,\bq^{\rm (0)})$.  The manifold, ${\cal R}^{(0)}$, formed by all such points is a reasonable zeroth-order approximation to the MEM $\cal R^*$. 

Set the integer value $P$ for the number of subspace relaxation iterations using the Newton-Raphson algorithm (see below), and validate it for the system. For example, $P=0$ (no relaxation) is the conventional CG-fixed rigidification approach, $P=1$ will give an exact result for a harmonic system, whereas $P=2$ will maintain the accuracy on the order of $1 \,{\rm cm}^{-1}$ for the CG free energy of $\HO_{10}$.

\subsection*{Algorithm to compute the coarse-grained potential }

\begin{enumerate}
    \item Start with a CG configuration \(\bR=(\bR_1,\dots,\bR_I)\) and initialize the associated point in the reference manifold, $\mathbf{r}^{(0)}\in {\cal R}^{(0)}$ by setting  $\br^{(0)}=(\bR,\bq^{\rm (0)})$ via \Eq{eq:bijection}.

    \item Iterate for \(p = 1, \dots, P\) using the Newton-Raphson method:
    \begin{equation} 
    \mathbf{r}^{(p)} = \mathbf{r}^{(p-1)} - \bM^{-1/2}\tilde{\bK}\bM^{-1/2}\ \nabla V(\br^{(p-1)})
    \end{equation}

     To a good approximation, the final configuration $\br^{(P)}\in {\cal R}^{(P)}$ lies near the MEM $\cal R^*$, i.e., ${\cal R}^{(P)}$ approximates $\cal R^*$.
    \item Compute the CG free energy $F(\bR;T)$  by setting $F(\bR;T)= F(\br^{(P)};T)$ and use either \Eq{eq:Fh} or \Eq{eq:Fh-classical}.
    
    \end{enumerate}

    For well-behaved systems, the pseudoinverse computed at the initial guess, $\tilde{\mathbf{K}}(\mathbf{r}^{(0)})$, often remains a good approximation throughout the iterations and hence $\tilde{\bK}$ need not be recomputed at each Newton-Raphson iteration (step 2 above). In the numerical example considered in this paper, we confirmed that this is the case.
However, a very high accuracy (thermodynamic consistency) requires the 
block-diagonal Hessian to be calculated again, at the final point $\br^{(P)}$, to obtain the normal 
mode frequencies used to evaluate the harmonic correction to the free energy (step 3 above).

\subsection*{Calculating observables}

Once the free energy, $F(\bR;T)$, is known as a function of the CG configuration $\bR$, observables depend only on $\bR$ can be computed straightforwardly using, e.g., Monte Carlo (MC) method. Paper I also provides expressions for estimating selected AA properties, such as the heat capacity and the intramolecular vibrational spectra. 
Moreover, it turns out that for no additional cost, an approximate ensemble of AA configurations $\br$ can be recovered during the simulation from the sampled CG configurations $\br^*$. The recovered distribution over $\br$ can be used to compute AA properties, in particular, spatial correlation functions or canonical averages of observables that depend explicitly on $\br$. This avoids the systematic bias introduced by evaluating AA observables directly on the MEM, where the absence of intramolecular fluctuations leads to distorted estimates for observables that depend on fast degrees of freedom.

To this end, for each block $i\in\{1,\dots,I\}$, update its internal coordinates by incorporating thermal fluctuations along its $L_i$ fast eigenmodes within the harmonic approximation:
\begin{equation}\label{eq:block_update}
\br_i = \br_i^* + \bM_i^{-1/2} \sum_{l=1}^{L_i} \xi_l^i\,\bU_l^i, 
\end{equation}
where $\xi_l^i\sim \mathcal{N} \left( 0, \, (\sigma_l^i)^2 \right)$ are independent Gaussian random variables with mean zero and the following variances:

\begin{equation}
(\sigma_l^i)^2 = 
\begin{cases}
\displaystyle \frac{\hbar}{2 \omega_l^i} \coth\left( \frac{\beta \hbar \omega_l^i}{2} \right) & \text{(Quantum)} \\
\displaystyle \frac{1}{\beta (\omega_l^i)^2} & \text{(Classical)}
\end{cases}
\end{equation}
A proper (i.e., consistent with the canonical ensemble) AA configuration is then obtained by reassembling the blocks, $\br = \left( \br_1,  \dots, \br_I \right)$.

\section*{Numerical example: Thermodynamic properties of the $\HO_{10}$ cluster}

In this section, we demonstrate that the equilibrium properties of a molecular cluster formed by 
relatively stiff molecules in a fully classical regime (i.e., not taking into account the nuclear 
quantum effects) can be computed both very efficiently and with near exact accuracy using the CG potential $F(\bR;T)$ (\textit{cf.} \Eq{eq:Fh-classical})
evaluated with the new SHR algorithm. 

We consider the $\HO_{10}$ cluster modeled by the AA q-TIP4P/F potential from Ref.~\onlinecite{habershon2009}.
All simulations were done by Replica Exchange Monte Carlo \cite{swendsen1986,geyer1991},
with $J=20$ replicas distributed in the temperature range [20-300]K using
the geometric temperature schedule, $T_j=T_{\rm min}(T_{\rm
  max}/T_{\rm min})^{j/(J-1)}$  ($j=0,...,J-1$). These simulations using the temperature-dependent CG potential and the calculation of observables followed the protocols described in Paper I.
To prevent cluster dissociation during the simulation, constraining sphere of radius $R_c =6.0\AA$ was used.
The SHR algorithm was implemented here using $P=2$ Newton-Raphson iterations, without updating the pseudoinverse $\tilde{\mathbf{K}}$ at each iteration, so that only one block-diagonal Hessian needed to be computed. At each MC step, the Hessian was computed for a second time, at the final point of the Newton-Raphson iteration, to evaluate the harmonic correction to the free energy.
We made sure that all calculations were well converged with respect to the total number of MC steps. Specifically, For the AA simulations, we used $~4.5\times 10^9$ MC steps, while for the CG simulations, $~2.2\times10^8$ MC steps. For the heat capacity, the statistical errors were well below 1 $k_{\rm B}$.

\begin{figure}[htbp]
	\begin{centering} 
	\includegraphics[width=0.9\linewidth]{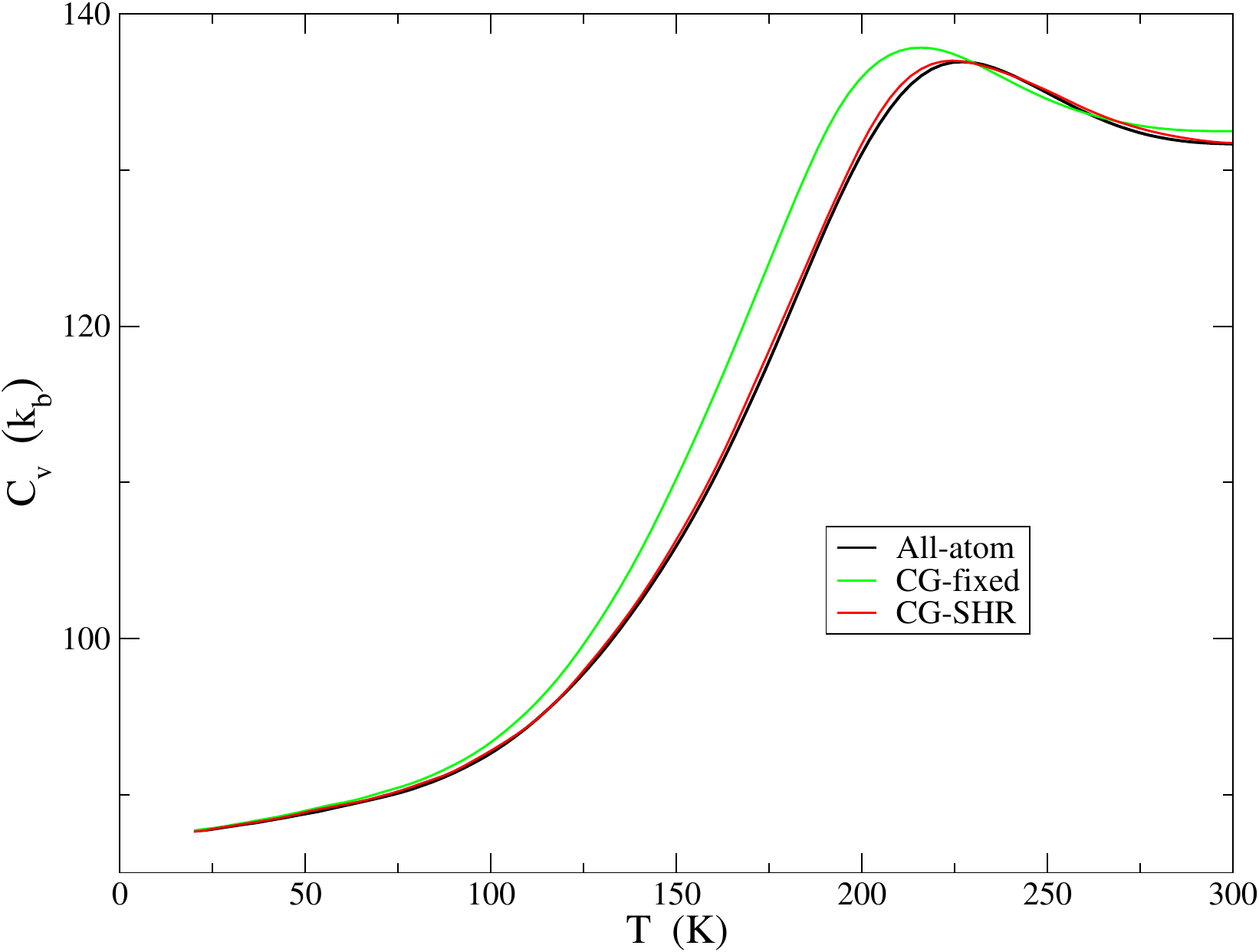}
	\end{centering}
	\caption{Heat capacity of the
          $\HO_{10}$ cluster computed by three different methods (see text). 
            \label{fig:Cv} } 
      \end{figure}

In \Fig{fig:Cv} we show the AA heat capacity $C_v(T)$ of
$\HO_{10}$ and compare it with the heat capacities computed using two different methods, namely: the conventional CG-fixed method, i.e., using the same fixed shape for all the molecular monomers, and the present CG-SHR method, i.e., Coarse-Graining with Subspace Harmonic Relaxation. For the latter, we found that using  $P=2$ Newton-Raphson iterations per MC step maintains very high accuracy. The agreement between the AA and CG-SHR results is within 1 $k_{\rm B}$, while the error for the CG-fixed approach is significant. Most importantly, the position of the maximum for the CG-fixed heat capacity is shifted to lower temperatures by about 10K.
Interestingly, compared with the AA simulation, the CG simulation required roughly an order of magnitude fewer MC steps to achieve a similar level of statistical error.

\begin{figure}[htbp]
	\begin{centering} 
	\includegraphics[width=0.9\linewidth]{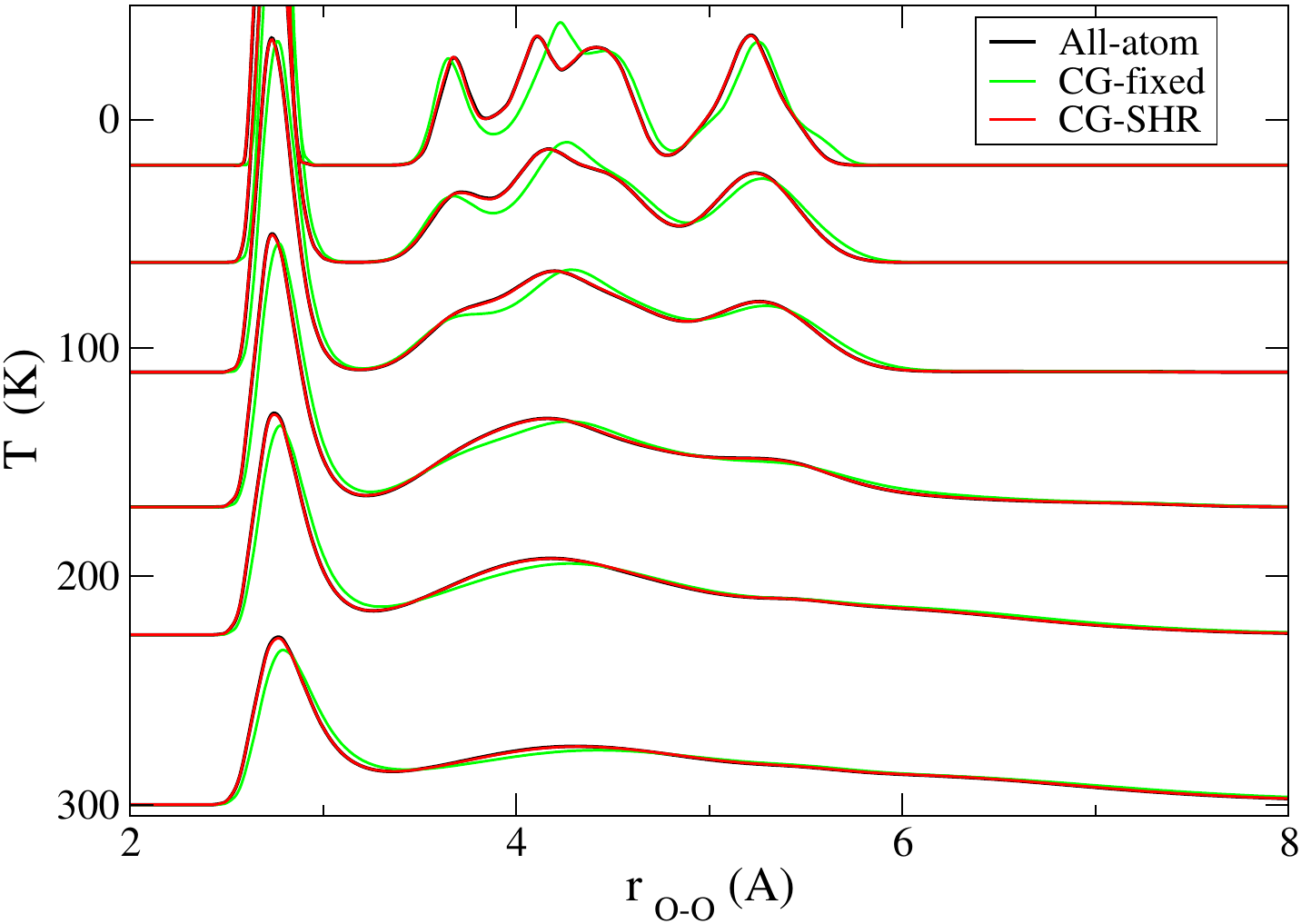}
	\end{centering}
	\caption{Same as \Fig{fig:Cv}, but for the O-O radial
          pair distribution functions at several selected temperatures. For presentation purposes, each distribution is shifted by the corresponding temperature value, which can be read at its crossing with the y-axis. The agreement between the AA and CG-SHR results is within the thickness of the lines. 
          \label{fig:OO} }
       \end{figure}

\Fig{fig:OO} compares the AA radial Oxygen-Oxygen distribution
functions for several selected temperatures with those computed using two different methods. While the distribution functions calculated by the present SHR method are essentially exact, the conventional CG-fixed calculations result in distributions that deviate significantly from the exact distributions, which is especially apparent at lower temperatures.

\begin{figure}[htbp]
	\begin{centering} 
	\includegraphics[width=0.9\linewidth]{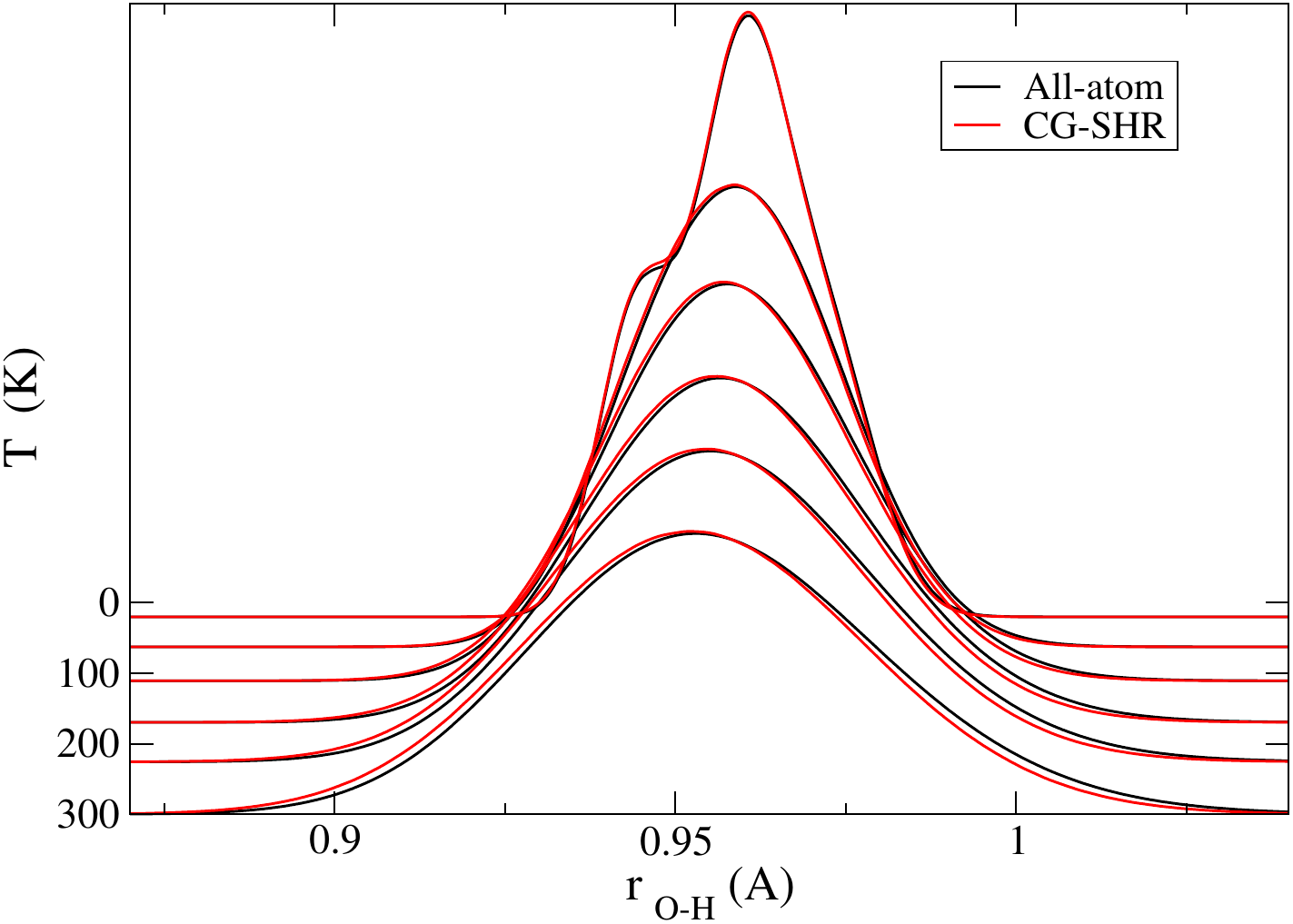}
	\end{centering}
	\caption{Same as \Fig{fig:OO} but for the O-H
          bond length distribution function, which is an AA property.
          \label{fig:OH} } 
       \end{figure}

\begin{figure}[htbp]
	\begin{centering} 
	\includegraphics[width=0.9\linewidth]{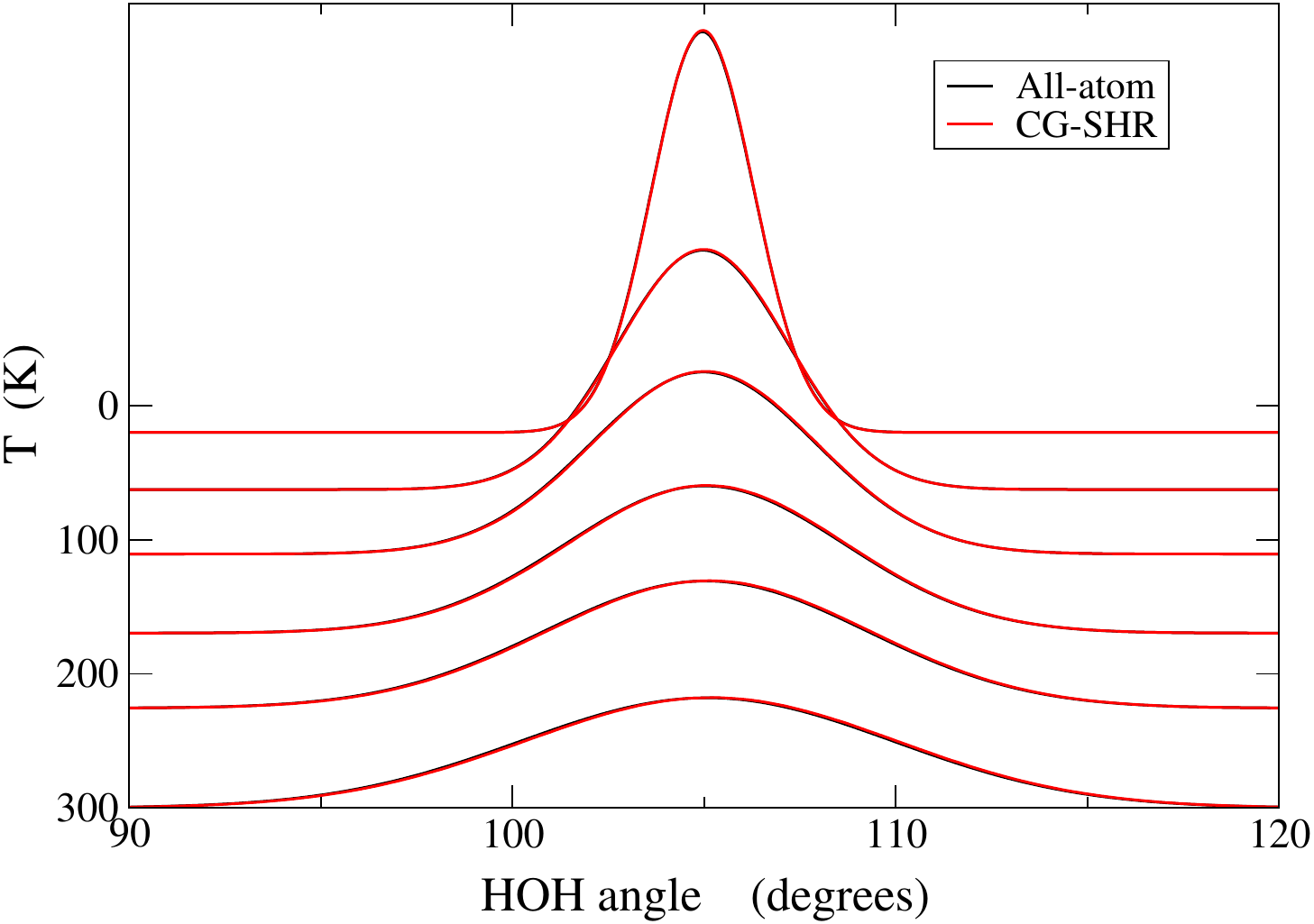}
	\end{centering}
	\caption{Same as \Fig{fig:OO} but for the H-O-H angle
          distribution function, which is an AA property.
          \label{fig:HOH} } 
       \end{figure}

Figures \ref{fig:OH} and \ref{fig:HOH} compare the OH bond length and HOH-angle distributions computed using the present SHR (applying \Eq{eq:block_update}) and the AA methods. Note that both correlation functions are AA properties, yet they are computed during the SHR simulation for negligibly small additional cost. The agreement between the exact (AA) and the present (SHR) methods is comparable to that for the heat capacity and the O-O pair distributions.
 
      \section*{Conclusions}

      In this paper, we revisited the recently introduced ``bottom-up'' CG method called here the ``Subspace Harmonic Relaxation''. Just like the conventional (CG-fixed) method of rigidification of molecular systems, the SHR method also computes the CG potential directly, while maintaining much higher accuracy. The CG-fixed method is based on fixing the configurations of the groups of atoms that are assumed to undergo only low-amplitude vibrations. In the SHR approach the fast degrees of freedom are forced to relax for each given CG configuration. This and the inclusion of harmonic corrections made the SHR potential much more accurate than the CG-fixed potential. In addition, in the context of solving the sampling problem, the SHR model benefits from a significant reduction of the dimensionality of the manifold to be sampled, an advantage over AA methods shared by all other CG models.
      
      However, the original formulation of the SHR method in Paper I could be significantly accelerated without compromising accuracy, in particular, because for the subspace minimization it used the conjugate gradient method, and even more importantly, because the normal mode calculation was done by diagonalizing the full Hessian matrix, which scales numerically as $\sim(3N)^3$. The present implementation of the SHR method is significantly more efficient than the original version proposed in Paper I. This was achieved here by replacing the full Hessian with its block-diagonal counterpart, and also by using the Newton-Raphson minimization instead of minimization by the conjugate gradient method.
Using the example of the $\HO_{10}$ cluster, we demonstrated that two evaluations of the block-diagonal Hessian and two gradient evaluations at each step of the Monte Carlo simulation are sufficient to maintain nearly exact thermodynamic consistency for the heat capacity and various spatial correlation functions. Remarkably, the present approach also allows one to compute all atom properties for very little additional cost. Due to the reduced number of MC steps required to achieve comparable statistical accuracy, the overall computational cost remains competitive with AA simulations.
      
      Finally, for large molecular systems, one may still find the improved algorithm for direct evaluation of the SHR potential too expensive. For such cases, the SHR model can benefit from its refitting using, for example, machine learning protocols designed for anisotropic shapes of rigid bodies and their interactions\cite{Campos-Villalobos2024}.
      
  \bigskip
   
   \section*{Acknowledgments}
   We would like to thank David Limmer,
      William Noid, David Wales, Florent Calvo, and Pinchen Xie for
      useful discussions.
      This work was supported by the National Science
      Foundation (NSF) Grant No. CHE-1900295.

    \section*{Data Availability Statement}
   	The data that supports the findings of this study are available within the article.

\bibliography{paper}

\end{document}